# Experimental and theoretical study of light scattering by individual mature red blood cells by use of scanning flow cytometry and discrete dipole approximation

**Maxim A. Yurkin, Konstantin A. Semyanov, Peter A. Tarasov,
Andrei V. Chernyshev, Alfons G. Hoekstra, and Valeri P. Maltsev**

*M. A. Yurkin, K. A. Semyanov, P. A. Tarasov, A. V. Chernyshev, and V. P. Maltsev are with the Institute of Chemical Kinetics and Combustion, Siberian Branch of the Russian Academy of Sciences, Institutskaya 3, Novosibirsk 630090 Russia*
*A. G. Hoekstra is with the Faculty of Science, Section Computational Science, of the University of Amsterdam, Kruislaan 403, 1098 SJ, Amsterdam, The Netherlands*
*M. A. Yurkin is also with the Faculty of Science, Section Computational Science, of the University of Amsterdam, Kruislaan 403, 1098 SJ, Amsterdam, The Netherlands*
*A. V. Chernyshev and V. P. Maltsev are also with Novosibirsk State University, Pirogova 2, Novosibirsk 630090 Russia*
*V. P. Maltsev's e-mail is maltsev@ns.kinetics.nsc.ru*

## Abstract

Elastic light scattering by mature red blood cells (RBCs) was theoretically and experimentally analyzed with the discrete dipole approximation (DDA) and the scanning flow cytometry (SFC), respectively. SFC permits measurement of angular dependence of light-scattering intensity (indicatrix) of single particles. A mature RBC is modeled as a biconcave disk in DDA simulations of light scattering. We have studied the effect of RBC orientation related to the direction of the incident light upon the indicatrix. Numerical calculations of indicatrices for several aspect ratios and volumes of RBC have been carried out. Comparison of the simulated indicatrices and indicatrices measured by SFC showed good agreement, validating the biconcave disk model for a mature RBC. We simulated the light-scattering output signals from the SFC with the DDA for RBCs modeled as a disk-sphere and as an oblate spheroid. The biconcave disk, the disk-sphere, and the oblate spheroid models have been compared for two orientations, i.e. face-on and rim-on incidence. Only the oblate spheroid model for rim-on incidence gives results similar to the rigorous biconcave disk model.

Keywords: red blood cell, discrete dipole approximation, non-spherical particle light scattering, scanning flow cytometry

*OCIS code:* 170.0170, 170.1530, 170.1470, 290.5850, 120.5820

## Introduction

The scattering of electromagnetic waves by dielectric particles is a problem of great significance for a variety of applications ranging from particle sizing and remote sensing to radar meteorology and biological sciences.[1,2] In the area of medical diagnostics, understanding how a laser beam interacts with blood suspensions or a whole-blood medium is of paramount importance in quantifying the inspection process in many commercial devices and experimental setups that are used widely for *in vivo* or *in vitro* blood measurements.[3] These measurements are focused mainly on electromagnetic scattering properties of red blood cells (RBCs), the type of cell that are most numerous in the blood.

Usually, RBCs measure from 6.6 to 7.5 μm in diameter, however, cells with a diameter greater than 9 μm (macrocytes) or less than 6 μm (microcytes) have been observed. These cells are nonnucleated (among vertebrates, only the red cells of mammalians lack nuclei) biconcave discs that are surrounded by thin, elastic membranes and filled with hemoglobin. They are soft, flexible, and elastic and therefore move easily through the narrow blood capillaries. The primary function of these cells is to carry oxygen from the lungs to the body cells.[3]

The light scattering properties of blood are based on a solution of the single-electromagnetic-scattering problem for a RBC. Moreover RBCs can play an important role in verification of solutions of the direct light-scattering problem for nonspherical particles because of their simple internal structure and stable biconcave discoid shape. Most of the relevant work solves the single-electromagnetic-scattering problem by a RBC analytically either basically by use of Mie[4], Fraunhofer[5] and anomalous diffraction theories[5,6] or numerically with the aid of the T-matrix approach[7] by treating a real RBC as a volume equivalent spherical[4], spheroidal,[7] or ellipsoidal[5,8] dielectric particle. In these cases the assumptions can be considered exact only because of the special experimental conditions when RBCs become spheroidal or ellipsoidal. In contrast, only a few papers have dealt with light scattering by a real nondeformed RBC.[9,10] Shvalov *et al.*[10] made use of the single- and double-wave Wentzel-Kramers-Brillouin approximations and were able to reproduce analytically the experimental results for forward scattering angles in the range of 15° to 35°. Mazeron and Muller,[11] with the aid of a physical optics approximation, presented small-angle forward-scattering patterns by a RBC whose axisymmetric geometry was obtained through complete rotation of a Cassini curve.

At present RBCs form the border of the size range where effective methods of light scattering simulation can be applied. Tsinopoulos and Polyzos[12] studied scattering by nondeformed, average-sized RBCs illuminated by a He-Ne laser beam. They examined various orientations of the RBC with respect to the direction of the incident light and computed scattering patterns in the forward, sideways, and backward directions. They used a boundary-element method appropriately combined with fast-Fourier-transform (FFT) algorithms.

Another method to simulate light scattering by arbitrary shaped particles is the discrete dipole approximation (DDA).[13] The latest improvements in the DDA[14] and its implementation on parallel supercomputers[15] permit simulation of light scattering by particles with the sizes of RBCs.

The next generation of flow cytometers, the scanning flow cytometer (SFC),[16] permits measurement of the angular dependence of the light-scattering intensity of single particles at a speed of $O(10^2)$ particles/s. In the research reported in this manuscript, light scattering by mature RBCs was theoretically and experimentally analyzed with the DDA and a SFC, respectively. In the numerical simulations we modeled a mature RBC as a biconcave disk.[17] The effect of RBC orientation relative to the direction of the incident light was studied. We carried out numerical calculations of indicatrices for several axis ratios and volumes of RBC. Comparison of the simulated indicatrices with those measured with the SFC showed a good agreement, validating the biconcave disk model for a mature RBC. We simulated light-scattering output signals from a SFC with DDA for a RBC modeled as a disk-sphere (a symmetric discoid layer cut out of a sphere) and an oblate spheroid and compared those results with results of biconcave disk model for two orientations, face-on and rim-on incidence.

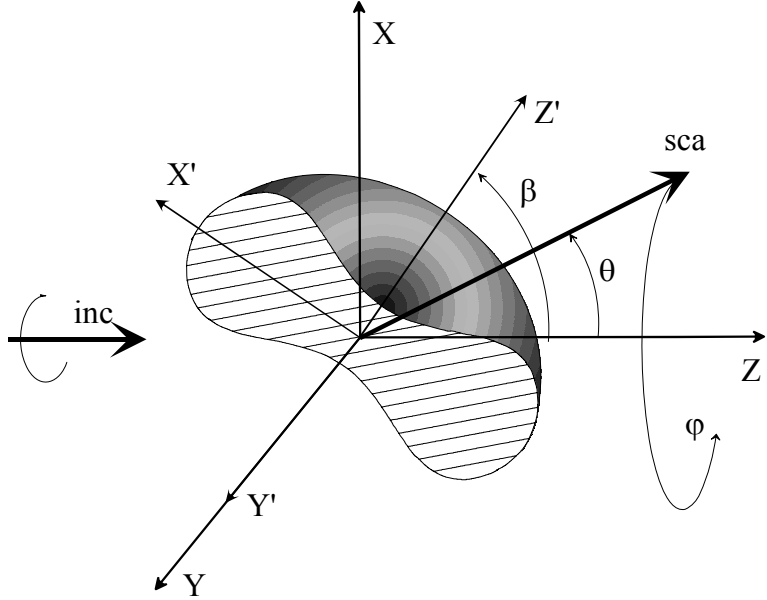

Fig. 1. Shape of a mature RBC and its orientation with respect to the incident radiation. *xyz* is a laboratory reference frame and *x'y'z'* is a reference frame tied to the RBC (*z'* is the axis of symmetry); inc and sca are propagation vectors for incident and scattering radiation respectively.

**Theory**

*Optical Model of the RBC*

A mature red blood cell (RBC) can be modeled as a biconcave discoid. A RBC is composed of hemoglobin (32%), water (65%), and membrane tissue (3%) and does not contain any nucleus.[18] The shape of the RBC was described by Fung *et al.*[17]:

$$T(x) = 0.65d\sqrt{1-x^2}\left(0.1583 + 1.5262x^2 - 0.8579x^4\right), \quad (1)$$

where $T$ is a thickness of RBC (along the axis of symmetry), $x$ is a relative radial cylindrical coordinate $x = 2\rho/d$ ($-1 \leq x \leq 1$), $\rho$ is a radial cylindrical coordinate, and $d$ is the diameter of the RBC. In order to vary the ratio of maximum thickness and diameter independently on the diameter of a RBC we have rewritten Eq. (1) in the following form:

$$T(x) = 2\varepsilon\, d\sqrt{1-x^2}\left(0.1583 + 1.5262x^2 - 0.8579x^4\right), \quad (2)$$

where $\varepsilon = T_{max}/d$ is an aspect ratio of maximum thickness and diameter. The RBC shape is shown schematically in Fig. 1.

*Discrete Dipole Approximation*

The DDA permits simulation of light scattering by a particle modeled as a finite array of polarizable points. The points acquire dipole moments in response to the local electric field. The dipoles interact by means of their electric fields,[19,20] and the DDA is also sometimes referred to as the coupled dipole approximation.[21] The theoretical base for the DDA, including radiative reaction corrections, was summarized by Draine and Flatau.[13] The performance of the DDA was significantly improved by introducing complex-conjugate gradient methods for solving systems of linear equations and a FFT algorithm for matrix-vector multiplication (summarized by Draine and Flatau[13]). Hoekstra et al.[15] parallelized the DDA and showed that the algorithm runs efficiently on distributed memory computers, provided that the number of dipoles per processor is large enough. They run DDA simulations with $O(10^7)$ dipoles, limited only by the available amount of memory.[15]

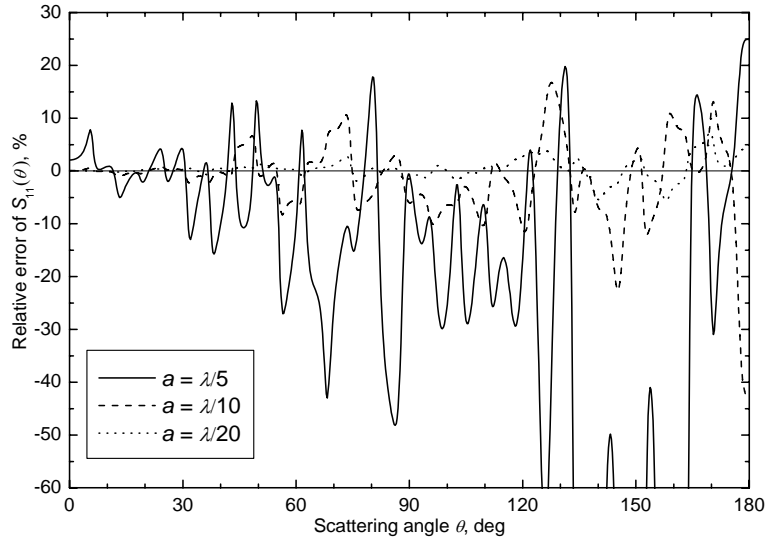

Fig. 2. Relative errors of DDA simulations for several discretization sizes. The DDA simulations were compared with the exact Mie solution for a sphere with diameter $d = 4.56$ μm and relative refractive index $m = 1.10$.

The size of a sub-volume that relates to the number of dipoles that form a particle must be small enough to ensure that response to an electromagnetic field is the response of an ideal induced dipole. The size should be in range $\lambda/20 < d < \lambda/10$.[13,22] In some cases we can even use $d \geq \lambda/10$. Comparison of light-scattering calculation performed with the DDA approximation and with exact Mie theory for a homogeneous sphere is a traditional way to determine the accuracy of the DDA. Using the parallel DDA code produced by Hoekstra et al.,[15] we compared DDA simulations with Mie calculations for a sphere with a diameter of 4.56 μm and a relative refractive index of 1.10; the wavelength was 0.6328 μm. The relative errors in DDA simulations as a function of the scattering angle are presented in Fig. 2 for three different discretization sizes: $\lambda/5$, $\lambda/10$, and $\lambda/20$. The relative error for discretization size of $\lambda/5$ is ~10% for scattering angles smaller then 50°. This degree of precision is good enough for comparison with experimental curves for which the scattered intensity was measured in a range of scattering angles from 10° to 50°. In the sequel of the paper, the size of the dipoles was varied in the range $\lambda/11$–$\lambda/8$ for different sizes of RBCs because of requirements for lattice regularity in the current implementation of the parallel FFT algorithm.[15]

## Experimental Equipment and Procedures

The experimental part of this study was carried out with a SFC that permits measurement of the angular dependency of light-scattering intensity (indicatrix) in a region ranging from 5° to 100°. The design and basic principles of the SFC were described in detail elsewhere.[16] The laser beam of the SFC is directed coaxially with the hydrodynamically focused flow which carries analyzed cells. Most disklike-shaped RBCs attain the testing zone of the SFC in rim-on orientation relative to the direction of the incident laser beam. The current set-up of the SFC provides a measurement of the following combination of Mueller matrix elements[23]:

$$I_s(\theta) = \int_0^{2\pi} d\varphi [S_{11}(\theta,\varphi) + S_{14}(\theta,\varphi)], \tag{3}$$

where $I_s(\theta)$ is the output signal of the SFC and $\theta$ and $\varphi$ are the polar and azimuthal angles, respectively. After integration over azimuthal angle $\varphi$ the second term in Eq. (3) vanishes, because of the axisymmetry of RBCs (see Eq. (A5) of Appendix A for details; we also obtained the same result numerically for all simulations). Therefore, the SFC output signal will be proportional to $S_{11}$ integrated over the azimuthal angle. To compare the experimental and theoretical light scattering from RBCs we used the DDA for calculation of the indicatrices. Our SFC setup allowed reliable measurements in the angular range 10°–50° (because of the operational range of the analog-digital converter).

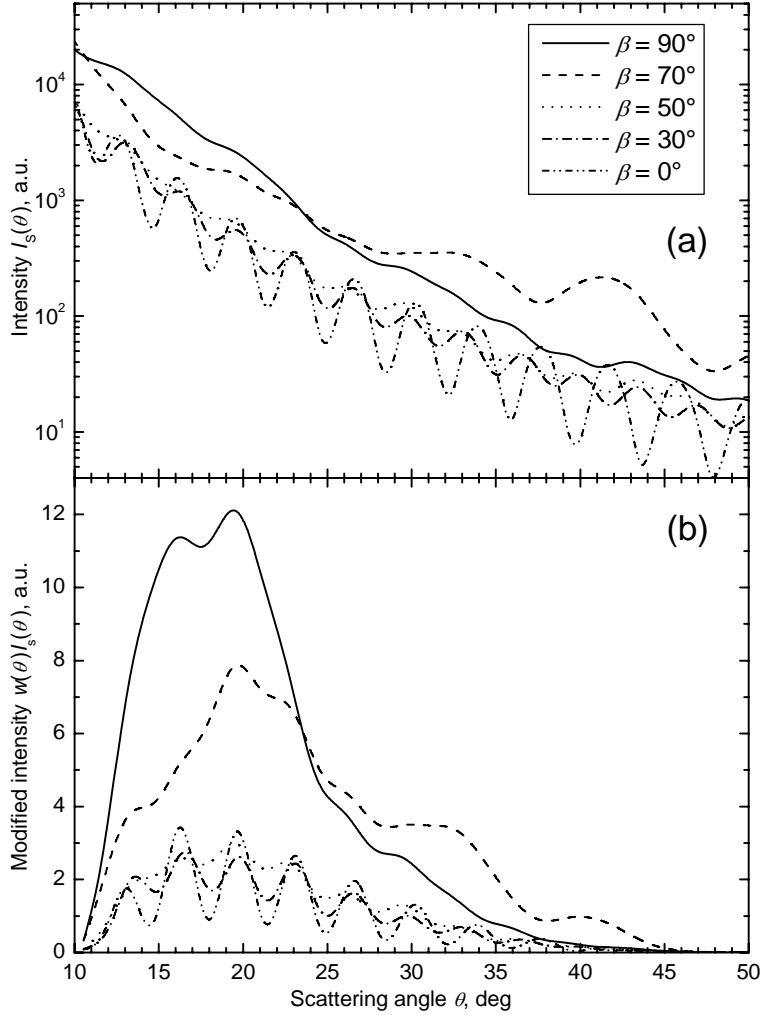

Fig. 3. (a) Initial and (b) modified indicatrices calculated by DDA for five orientations of the single biconcave disk relative to the direction of incident light beam.

A sample containing approximately $10^6$ cells/ml was prepared from fresh blood with buffered saline used for dilution. We used the SFC to continuously measure 3000 indicatrices of RBCs. Each of them was compared with each of the calculated theoretical indicatrices (for details see next section) by a $\chi^2$ test with a weighting function:

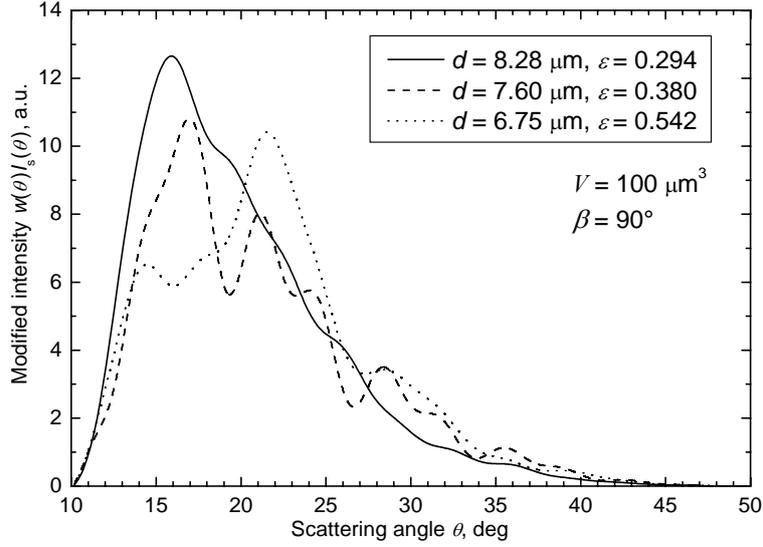

Fig. 4. Modified indicatrices of biconcave disks with different diameters and fixed volume.

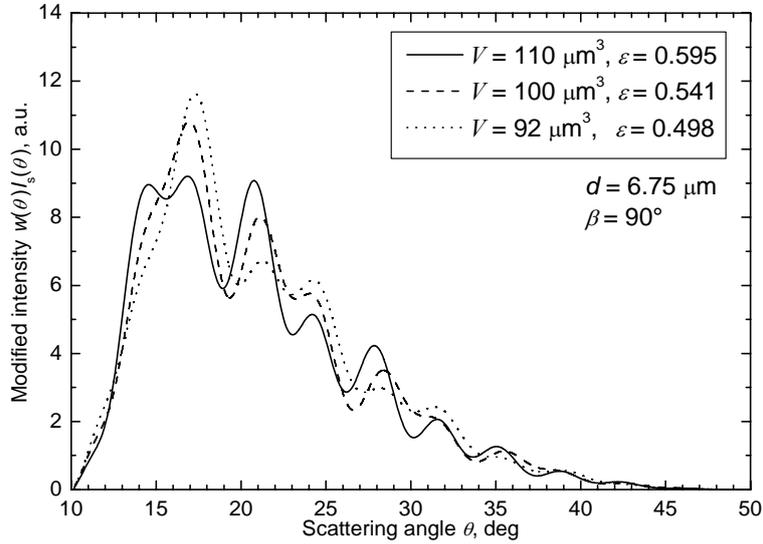

Fig. 5. Modified indicatrices of biconcave disks with different volumes and fixed diameter.

$$\chi^2 = \frac{\sum_{i=1}^{N}[I_{\exp}(\theta_i) - I_{\text{theor}}(\theta_i)]^2 w^2(\theta_i)}{N}, \quad (4)$$

where $I_{\exp}(\theta_i)$ is the experimental indicatrix measured at $N$ angular values $\theta_i$ ($\theta_1 = 10°$, $\theta_N = 50°$), $I_{\text{theor}}(\theta_i)$ is the theoretical indicatrix calculated at the same angles, $w(\theta_i)$ is a weighting function, defined as:

$$w(\theta_i) = \sin^2\left(\pi \frac{\theta_i - 10°}{50° - 10°}\right), \quad (5)$$

which is intended to suppress experimental errors (a detailed discussion is presented in Ref. 24). An experimentally measured RBC is said to have characteristics of the closest (by $\chi^2$) theoretical indicatrix, if their $\chi^2$ distance is less than a threshold. The threshold was set empirically to 40.

The optical model of the RBC was used in calculations with the profile described by Eq. (2). The RBC diameter was varied from 6 to 9 μm. We assume that the imaginary part of the refractive index is negligible at the wavelength $\lambda$ of 0.6328 μm. The refractive index of the surrounding medium (saline) was 1.333. The RBC refractive index was fixed at 1.40 that falls into the region of typical variation of a RBC refractive index (1.39 – 1.42)[25] and results in a relative refractive index of 1.05.

## Results and discussions

Using the DDA, we studied the effect of the biconcave disk orientation on the indicatrices. Angular orientation $\beta$ is the angle between the axis of the biconcave disk symmetry and the direction of the incident beam (Fig. 1). We used the profile defined by Eq. (2) with the following typical RBC characteristics[17]: diameter of 8.28 µm, aspect ratio of 0.323 (resulting in a volume of 110 µm$^3$), and relative refractive index of 1.05. The indicatrices for five orientations are shown in Fig. 3(a). To provide an effective comparison of experimental and theoretical light-scattering data we modified the indicatrices by multiplication by weighting function $w(\theta_i)$ [Eq. (5)]. The multiplication corresponds to the standard Hanning window procedure that strongly reduces the effects of the discontinuities at the beginning and the end of the sampling period of the SFC.[24] Moreover, this function resembles the SFC instrument function[16] and improves visual inspection of indicatrices, as the logarithmic scale can be replaced by a linear one. The modified indicatrices for the five orientations are shown in Fig. 3(b). The orientation of the biconcave disk modifies the indicatrix substantially. For instance, the mean scattering intensity, in the angular range from 15° to 50°, has a reduced intensity for symmetric orientation ($\beta = 0°$), whereas the oscillating structure vanishes for increasing orientation angle. Therefore the mean light-scattering intensity and indicatrix structure are good indicators of a mature RBC orientation in scanning flow cytometry.

Practically, it is interesting to study the sensitivity of indicatrix structure to variations in the volume and diameter of the biconcave disk when the orientation corresponds to orthogonal directions of the biconcave disk symmetry axis and incident beam (i.e. for $\beta = 90°$). The last requirement is caused by the orientating effect of the hydrofocussing head of the SFC in which nonspherical particles are oriented with a long axis along the flow lines.[9] We calculated the indicatrices of RBC with a typical volume of 100 µm$^3$ while we varied the biconcave disk diameter. The results of the DDA simulation are shown in Fig. 4. Variation of the diameter does not change the intensity substantially, whereas the visibility of the oscillating structure, which characterizes the relative difference between maximum and minimum values,[16] is reduced with increasing biconcave disk diameter.

Additionally, we varied the thickness of the biconcave disk, which resulted in variation of volume with a constant diameter. The RBC volume is *the* most important hematological index measured with modern automatic hematology analyzers.[3] Flow cytometry[25] and Coulter's cell[3] are the instrumental solutions utilized for measurement of RBC volume distribution. The modified indicatrices of biconcave discs calculated for three volumes with a typical diameter of 6.75 µm are shown in Fig. 5. The light-scattering intensity does not depend on variation of the biconcave disk volume or thickness. There is the same tendency in the indicatrix structure as for Fig. 4: the visibility of the oscillating structure increases for increasing aspect ratio (of thickness and diameter).

Our simulation of light scattering by mature RBCs allows us to conclude that the orthogonal orientation of a biconcave disk relative to the direction of incident beam does not provide sufficient sensitivity of the indicatrix to the RBC characteristics. Coincidence of biconcave disk symmetry axis and direction of incident beam (i.e. $\beta = 0°$) gives an advantage in the solution of the inverse light-scattering problem because in that case the oscillating structure of the indicatrix is much more sensitive to variation of RBC characteristics. Unfortunately, as it is shown below, such orientation is improbable in SFC.

Table 1. The parameters of RBCs for preliminary calculations.[a]

| $d$, μm | $V$, μm³ | | | | |
|---|---|---|---|---|---|
| | 86 | 92 | 100 | 105 | 110 |
| 6.08 | 0.638 | – | – | – | – |
| 6.33 | 0.565 | 0.605 | – | – | – |
| 6.51 | – | 0.556 | 0.604 | – | 0.665 |
| 6.75 | 0.466 | 0.499 | 0.542 | 0.569 | 0.596 |
| 6.84 | – | – | 0.521 | – | 0.573 |
| 7.01 | – | – | 0.484 | – | 0.532 |
| 7.60 | 0.327 | 0.349 | 0.380 | 0.399 | 0.418 |
| 8.28 | – | – | 0.294 | 0.418 | 0.323 |

[a] Values are aspect ratios for selected diameter and volume. Dashes indicate that indicatrices were not calculated.

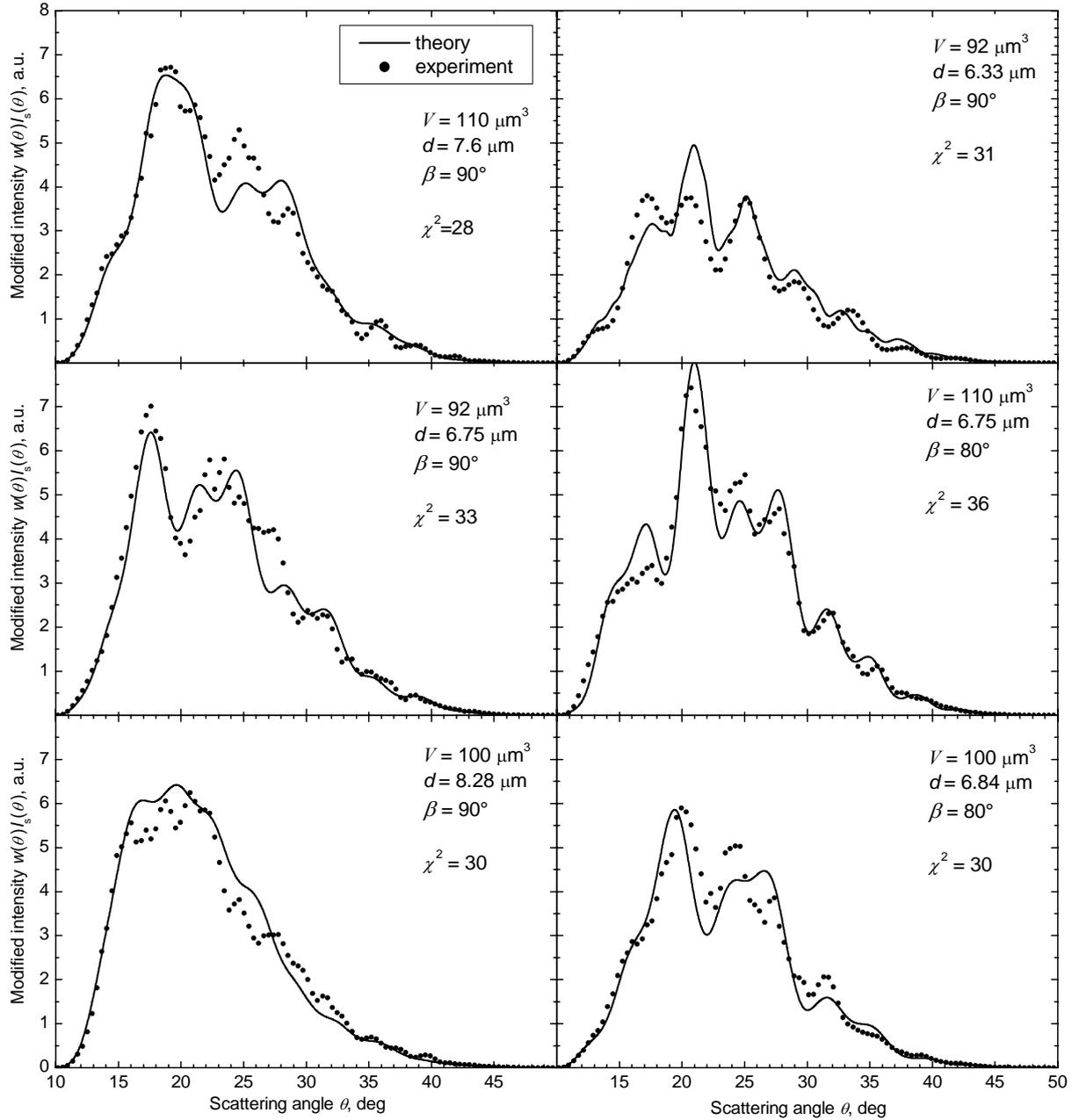

Fig. 6. Experimental and theoretical modified indicatrices of individual mature RBCs. Values of $\chi^2$ differences are shown.

Practically, the DDA cannot be used to fit experimental indicatrices because the DDA calculations require approximately 10 h for a fixed biconcave disk orientation (the computations where carried out on 8 nodes of a Beowulf computer with nodes running at 750 MHz). To solve this problem we made calculations of indicatrices for several biconcave disks with different diameters and volumes, presented in Table 1, to fill a small database to be used in the inverse problem. Values in the cells as listed in the table are aspect ratios for selected diameter and volume, whereas dashes indicate that indicatrices of RBC with these characteristics were not calculated. For each set of diameter and volume, four indicatrices were calculated for orientation angles $\beta$ (= 60°, 70°, 80°, 90°). Calculated theoretical indicatrices were used for determination of characteristics of experimentally measured RBCs by the $\chi^2$ test as explained in the previous section.

A few representative results of our comparison by the $\chi^2$ test are presented in Fig. 6. One can see that theoretical indicatrices fit experimental curves well. The mean scattering intensity (intensity integrated over the angular interval) of the measured indicatrices is in a good agreement with the mean intensity of theoretical indicatrices calculated for orientation angles $\beta$ of 70° and 90° [Fig. 3(b)], and exceeds the mean intensity for smaller orientation angles. This fact allows us to conclude that orientation of RBCs in the capillary of a SFC is close to orthogonal ($\beta = 90°$), which agrees with our previous results.[9] This result justifies our choice of orientation angles for preliminary calculations of RBCs' indicatrices. We used all the experimental indicatrices that passed the $\chi^2$ test to plot a distribution of mature RBCs over the orientation angle, as presented in Fig. 7. This distribution proves once again that orthogonal orientation is much preferable for RBCs in the capillary of a SFC and gives an estimation of deviation from orthogonal orientation.

Contrary to the DDA, the T-matrix method allows one to reduce a time of light-scattering calculation substantially.[2] However the biconcave disk shape complicates the light-scattering simulation in the T-matrix method. This method can be effectively applied to a particle with disk-sphere or oblate spheroid shapes that can be used as models of RBCs. Therefore, we have compared the indicatrices of particles shaped according to Eq. (2) and of particles with the following shape geometries: disk-sphere and oblate spheroid. The comparison was performed for the diameter-volume-equal particles. The parameters of the biconcave disk were as follows: $d = 7.60$ µm, $\varepsilon = 0.380$ ($V = 100$ µm$^3$), and $m = 1.05$. The light scattering for two orientations of the particles relative to the direction of the incident beam, rim-on and face-on incidence, was computed. The indicatrices of the biconcave disk and disk-sphere are shown in Fig. 8. This figure allows us to conclude that RBC can be modeled by a disk-sphere only for simulation of light scattering in the angular interval ranging from 10° to 15°.

An oblate spheroid is the most popular model in simulation of light scattering of individual RBCs. This model was used in simulation of light scattering by the T-matrix method.[7] We have calculated the indicatrices of diameter-volume-equal spheroids in rim-on and face-on orientations of the oblate spheroid and RBC. The results of these calculations, shown in Fig. 9, allow us to conclude that RBC can be modeled by an oblate spheroid over a wide angular interval only for rim-on incidence. This conclusion is in agreement with the boundary-element methodology[12] and discrete sources method[26] applied to study of light scattering of red blood cell. However the validity of such substitution for calculation of RBC indicatrices should be further studied for different RBC sizes with respect to the certain problem, where these indicatrices are to be used.

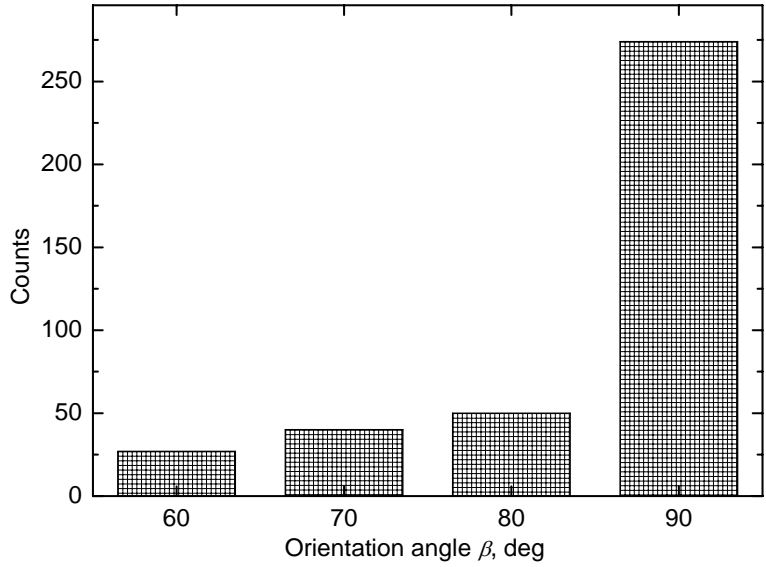

Fig. 7. Distribution of mature RBCs over an orientation angle obtained using $\chi^2$ test.

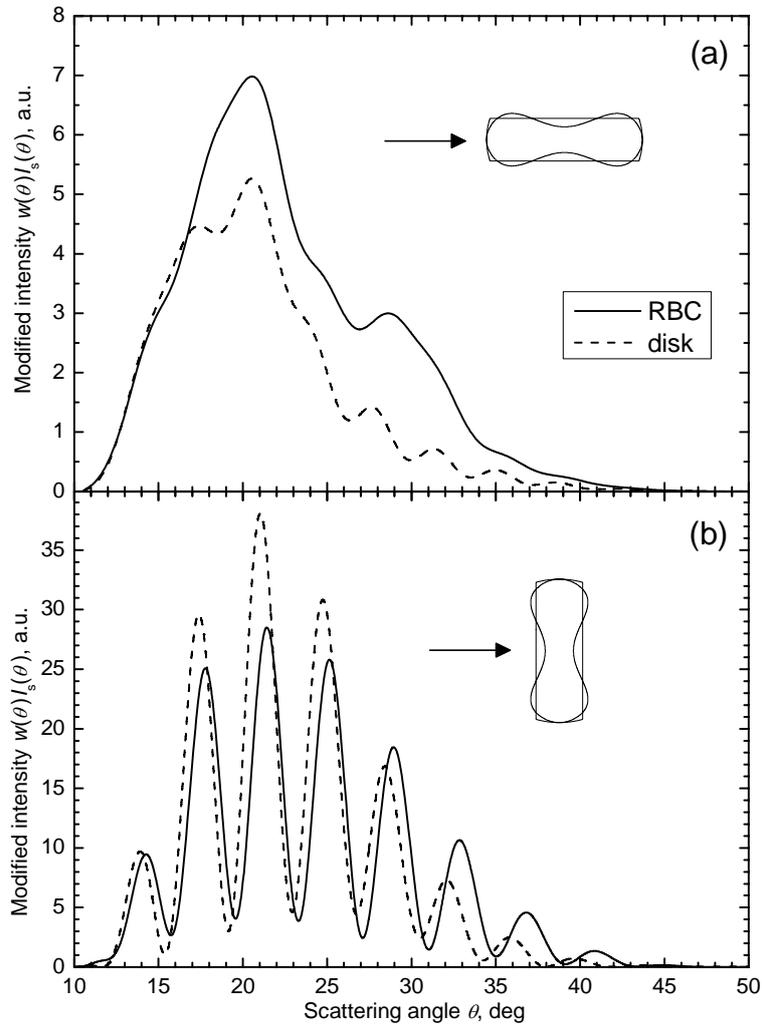

Fig. 8. The modified indicatrices of the biconcave disk and the diameter-volume-equivalent disk-sphere: (a) rim-on and (b) face-on incidence.

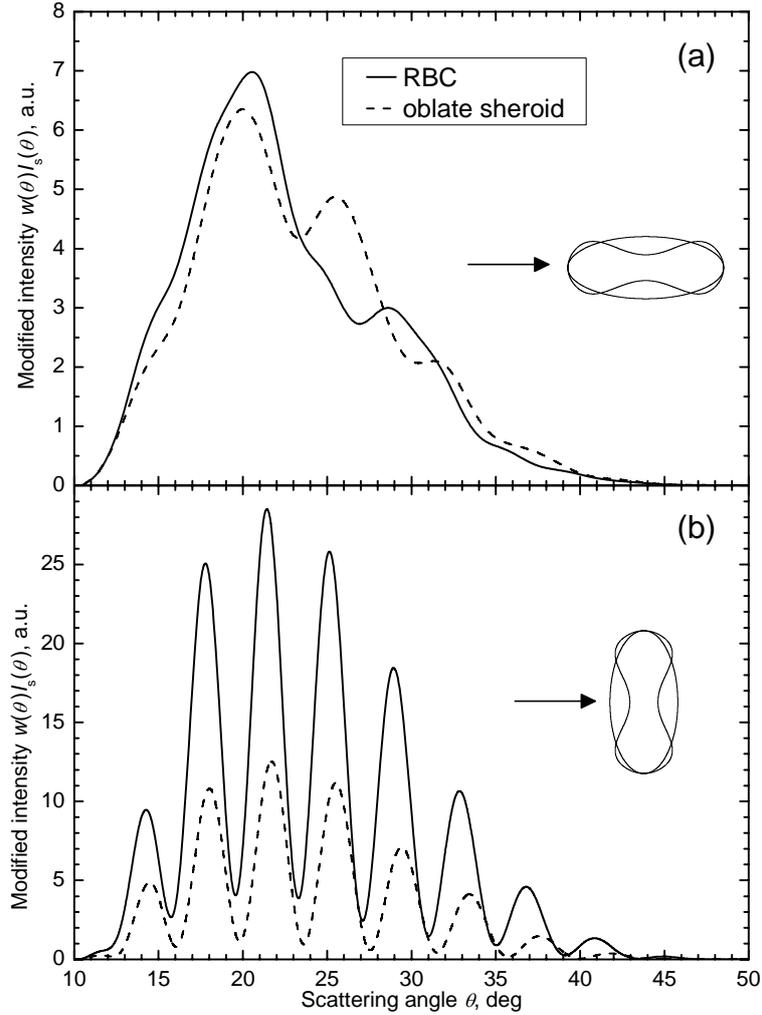

Fig. 9. The modified indicatrices of the biconcave disk and the diameter-volume-equivalent oblate spheroid: (a) rim-on and (b) face-on incidence.

## Conclusion

Our simulation of light scattering of a mature RBC has shown that the indicatrix is sensitive to the RBC shape and the DDA (or some other method for which no simplifications of the RBC shape are assumed) should be used in a study of formation of the indicatrix for RBCs with various characteristics. However, light scattering of RBCs can be simulated with the T-matrix method for rim-on incidence by use of the oblate spheroid model. Fortunately, the hydrodynamic system of the SFC delivers mature RBCs into the testing zone in this specific orientation. This performance of the SFC gives a chance of solving the inverse light-scattering problem for mature RBCs, e.g. by parameterization or use of a neural network because T-matrix simulation requires substantially less computing time than the DDA algorithm. However, the precision of such algorithms when developed should be tested using realistic indicatrices obtained e.g. by DDA simulations.

Therefore, need for improvement of current DDA code arises. Such an improvement can not only provide enough testing indicatrices for the problem described above but also make feasible such tasks as solving an inverse light-scattering problem for any orientation of RBCs relative to the incident beam.

This research was supported by Russian Foundation for Basic Research through the grant 02-02-08120-inno and 03-04-48852-a, by Siberian Branch of the Russian Academy of Sciences through the grant 115-2003-03-06, and by the NATO Science for Peace program through grant SfP 977976.

## Appendix A

Let incident radiation propagate along the $z$-axis, and assume that a particle has a symmetry plane containing this axis. We will investigate then the properties of the Mueller matrix integrated over complete azimuthal angle $\varphi$ at a fixed polar angle $\theta$.

Not restricting generality (as we consider the integral over the complete azimuthal angle), we can consider that the $x$-axis resides in the symmetry plane of the particle. We divide the integral into two parts and then group them:

$$\int_{-\pi}^{\pi} d\varphi \mathbf{S}(\theta,\varphi) = \int_{-\pi}^{0} d\varphi \mathbf{S}(\theta,\varphi) + \int_{0}^{\pi} d\varphi \mathbf{S}(\theta,\varphi) = \int_{0}^{\pi} d\varphi [\mathbf{S}(\theta,\varphi) + \mathbf{S}(\theta,-\varphi)]. \tag{A1}$$

Let us consider two scattering problems for angles $(\theta,\varphi)$ and $(\theta,-\varphi)$. We rotate the laboratory reference frame about the $z$-axis by angles $\varphi$ and $-\varphi$ for the first and the second problems respectively (it is the same as rotating everything else – the particle, the incident and scattering direction, and the electric field vectors – in the backward direction). These operations do not change the scattering matrices, therefore

$$\mathbf{S}(\theta,\varphi) = \mathbf{S}_{-\varphi}(\theta,0) \text{ and } \mathbf{S}(\theta,-\varphi) = \mathbf{S}_{\varphi}(\theta,0), \tag{A2}$$

where $\mathbf{S}_\varphi$, $\mathbf{S}_{-\varphi}$ are scattering matrices for particles ($p_\varphi$ and $p_{-\varphi}$) rotated about the $z$-axis by angle $\varphi$ and $-\varphi$ respectively, relative to its initial orientation ($p_0$). Let us denote operator of plane $zx$ reflection as $P_{zx}$ and operator of rotation around the axis $z$ by angle $\varphi$ as $R_\varphi$. Then

$$P_{zx} p_{-\varphi} = (R_\varphi \circ P_{zx} \circ R_\varphi) p_{-\varphi} = (R_\varphi \circ P_{zx}) p_0 = R_\varphi p_0 = p_\varphi, \tag{A3}$$

where the first equation is an identity for any operand (which can be easily verified since operators involved do not affect $z$ coordinates) and the third one exploits the assumption that particle has a $zx$ symmetry plane in its initial orientation. Equations (A2) and (A3) imply that $\mathbf{S}(\theta,\varphi) + \mathbf{S}(\theta,-\varphi)$ can be considered as a sum of scattering matrices for the same scattering geometry but for particles which are mirror images of each other with respect to the scattering plane. It was shown[6] that such a sum gives a scattering matrix of the form:

$$\begin{pmatrix} S_{11} & S_{12} & 0 & 0 \\ S_{21} & S_{22} & 0 & 0 \\ 0 & 0 & S_{33} & S_{34} \\ 0 & 0 & S_{43} & S_{44} \end{pmatrix}. \tag{A4}$$

Then Eq. (A1) obviously implies that the Mueller matrix integrated over the complete azimuthal angle will be of the same form. Therefore we have proved that, if particle has a symmetry plane containing the $z$-axis (coincident with the propagation vector of the incident radiation), then

$$\oint d\varphi S_{ij}(\varphi) = 0, \text{ for } i = 1, 2 \text{ and } j = 3, 4 \text{ or vice versa}. \tag{A5}$$

It is clear that, for a body of rotation, any plane containing symmetry axis is as well a symmetry plane. Therefore a plane containing both the symmetry axis and the axis $z$ is symmetric for the body of rotation regardless of its orientation. Hence Eq. (A5) is automatically proved for axisymmetric particles.